# Blue photoluminescence from chemically derived graphene oxide

Goki Eda[1], Yun-Yue Lin[3], Cecilia Mattevi[1], Hisato Yamaguchi[2], Hsin-An Chen[3], I-Sheng Chen[3], Chun-Wei Chen[3*], and Manish Chhowalla[1,2†]

[1] *Department of Materials, Imperial College, Exhibition Road, London SW7 2AZ, UK.*
[2] *Department of Materials Science and Engineering, Rutgers University
607 Taylor Road, Piscataway, NJ 08854, USA.*
[3] *Department of Materials Science and Engineering, National Taiwan University
No. 1, Sec. 4, Roosevelt Road, Taipei 10617, Taiwan.*

Fluorescent organic compounds are of significant importance to the development of low-cost opto-electronic devices[1]. Blue fluorescence from aromatic or olefinic molecules and their derivatives is particularly important for display and lighting applications[2]. Thin film deposition of low-molecular-weight fluorescent organic compounds typically requires costly vacuum evaporation systems. On the other hand, solution-processable polymeric counterparts generally luminesce at longer wavelengths due to larger delocalization in the chain. Blue light emission from solution-processed materials is therefore of unique technological significance. Here we report near-UV to blue photoluminescence (PL) from solution-processed graphene oxide (GO). The characteristics of the PL and its dependence on the reduction of GO indicates that it originates from the recombination of electron-hole (e-h) pairs localized within small $sp^2$ carbon clusters embedded within an $sp^3$ matrix. These results suggest that a sheet of graphene provides a parent structure on which fluorescent components can be chemically engineered without losing the macroscopic structural integrity. Our findings offer a unique route towards solution-processable opto-electronics devices with graphene.

[†]E-mail: m.chhowalla@imperial.ac.uk
[*]E-mail: chunwei@ntu.edu.tw

Graphene is an exciting material for fundamental and applied solid state physics research[3]. Novel condensed matter effects arising from its unique two dimensional energy dispersion along with superior properties make it potentially useful for a wide variety of applications. However, graphene is a zero band gap semiconductor, which creates a unique set of challenges for implementation into conventional electronics due to substantial leakage currents in field effect devices[4]. The lack of a band gap also makes the possibility of observing luminescence highly unlikely.

Recently, chemically derived GO has been receiving attention for large-area electronics because its solubility in a variety of solvents allows ease of wafer-scale deposition[5-11]. Transport of carriers in reduced GO is limited by the structural disorder[12, 13]. However, conductivity of ~ $10^5$ S/m (Ref. [5, 6, 14]) and mobilities of ~ 10 $cm^2$/V-s (Ref. [15]) are sufficiently large for applications where inexpensive and moderate performance electronics (such as on flexible platforms) are required. In GO, large fraction (0.5 ~ 0.6) of carbon is $sp^3$ hybridized and is covalently bonded with oxygen in a form of epoxy and hydroxyl groups[16, 17]. The remaining carbon is $sp^2$ hybridized and are bonded either with neighboring carbon atoms or with oxygen in the form of carboxyl and carbonyl groups, which predominantly decorate the edges of the graphene sheets. GO is therefore a two dimensional network of $sp^2$ and $sp^3$ bonded atoms, in contrast to an ideal graphene sheet which consists of 100 % $sp^2$ hybridized carbon atoms. This unique atomic and electronic structure of GO[18], consisting of variable $sp^2$/$sp^3$ fractions, opens up possibilities for new functionalities. The most notable difference between GO and mechanically exfoliated graphene is the opto-electronic properties arising from the presence of a finite band gap[19]. Recently, PL from chemically derived GO has been demonstrated[20-22]. The luminescence of GO was found to occur in the visible and near infrared (IR) wavelengths range, making it useful for bio-sensing and identification tags[20, 21].

In carbon materials containing a mixture of $sp^2$ and $sp^3$ bonding, the opto-electronic properties are determined by the π states of the $sp^2$ sites[23]. The π and π* electronic levels of the $sp^2$ clusters lie within the band gap of σ and σ* states of the $sp^3$ matrix and are strongly localized[24, 25]. The optical properties of disordered carbon thin films containing a mixture of $sp^2$ and $sp^3$ carbon have been widely investigated[26-30]. The PL in such carbon systems is a consequence of geminate recombination of localized e-h pairs in $sp^2$ clusters which essentially behave as the luminescence centers, or chromophores[31]. Since the band gap depends on the size, shape, and fraction of the $sp^2$ domains, tunable PL emission can be achieved by controlling the nature of $sp^2$ sites. For example, PL energy linearly scales with the $sp^2$ fraction in disordered carbon systems[32].

Here we demonstrate that moderately reduced GO thin films consisting of several mono-layers emit near-UV blue light when excited with UV radiation. The blue PL was observed for thin film samples deposited from thoroughly exfoliated suspensions. We also observed red and near IR emission, comparable to Ref. [20-22] (see Supplementary Information), on GO films drop-casted from poorly dispersed suspensions. We demonstrate that by appropriately controlling the concentration of isolated $sp^2$ clusters through reduction treatment, the PL intensity can be increased by a factor of 10 as compared to the as-synthesized material. Reduction of GO can be achieved in a number of ways to obtain graphene-like electrically conductive material[11]. A common reduction method is exposure to hydrazine vapor, which leads to transformation of GO from an insulator to a semimetal[7]. The gradual transformation of GO is confirmed by absorbance measurements shown in Figure 1a on the same film at various stages of reduction (i.e. hydrazine vapor exposure time). The main absorbance peak attributed to $\pi - \pi^*$ transitions of C=C in as-synthesized GO occurs around ~ 200 nm which red shifts to ~ 260 nm upon reduction. The broad absorption spectra that extend up to 1500 nm indicate the absence of a well-defined band edge in the UV-visible energy range. A shoulder around 320 nm observed for as-synthesized GO may be attributed to $n - \pi^*$ transitions of C=O (Ref. [33]). This shoulder disappears almost immediately after exposure to hydrazine treatment (see Supplementary Information), most likely due to the decrease in concentration of carboxyl groups. The absorbance is found to increase with hydrazine exposure time, consistent with the evolution of oxygen (from ~ 39 at. % in starting GO to 7 – 8 at.% in the reduced GO) and concomitant increase in the $sp^2$ fraction from 0.4 to 0.8 (Ref. [14, 17]).

The corresponding PL spectra of the GO thin film after each incremental hydrazine vapor exposure (from 20 seconds up to 60 minutes) are shown in Figure 1b. It is immediately clear that in contrast to the broad absorption features, relatively narrow PL peak (FWHM ~ 0.6 eV) centered around 390 nm is observed. The liquid suspensions of GO used for the film deposition exhibited an apparently equivalent PL peak centered around 440 nm (See Supplementary Information). The shift in the suspension PL can be attributed to the difference in the dielectric environment[34]. The PL intensity was weak but was visible by unaided eyes for the excitation conditions used in this study. Preliminary measurements have indicated that the quantum yield is also very low but a detailed study of the quantum efficiency will be presented elsewhere. The PL signal, however, was sufficiently strong to allow reproducible data collection using the described measurement conditions.

The PL peak positions of the thin films were found to remain constant around ~ 390 nm with reduction treatment, varying by less than ± 10 nm. It can be seen that while the PL intensity is weak for as-deposited GO films, short exposure to hydrazine vapor results in dramatic increase in the PL intensities. Interestingly, this trend is reversed after > 3 min exposure to hydrazine vapor. Longer exposure leads to eventual quenching of the PL signal. It should be mentioned that reduction by thermal annealing at 200 °C in vacuum also led to weaker PL signal (See Supplementary Information). The PL excitation spectra at different PL emission wavelengths for a GO film reduced for 3 minutes are shown in Figure 1c. Excitonic features are readily observable between excitation wavelengths of 260 and 310 nm (4 ~ 4.4 eV), which represent the absorption energies corresponding to emission of blue light. The general features of the PL emission/excitation spectra and their dependence on the degree of reduction are markedly different from the red-IR emission reported in previous studies[20-22], suggesting that the origin of the blue emission is also different. It should be noted that for the observation of UV-blue PL, it is necessary to minimize the concentration of multi-layered and aggregated flakes via centrifugation. Since we observed red-near-IR PL from GO suspensions prior to centrifugation, we suspect that low-energy PL can be attributed to the presence of such particles in which inter-layer electronic relaxation can occur.

One possible origin of the blue PL is the radiative recombination of electron-hole pairs generated within localized states. The energy gap between the π and π* states generally depends on the size of $sp^2$ clusters[23] or conjugation length[35]. From Raman and imaging analysis, it has been suggested that GO consists of ~ 3 nm $sp^2$ clusters isolated within $sp^3$ carbon matrix[12, 14, 18]. Although no direct observation of "molecular" $sp^2$ domains has been reported, our transport studies in progressively reduced GO[14, 36] and previous works on PL from amorphous carbons[28, 37, 38] suggest their presence. It is the interactions between the nanometer size $sp^2$ clusters and the finite sized "molecular" $sp^2$ domains that is the key in optimizing the blue emission in GO. The $sp^2$ clusters having a diameter of ~ 3 nm consist of > 100 aromatic rings. Our calculations based on Gaussian and time-dependent (TD) DFT (See Supplementary Information for details) indicate that such $sp^2$ clusters have energy gaps of around 0.5 eV and cannot be responsible for the blue emission observed here. Figure 2 shows that the calculated HOMO-LUMO gap of a single benzene ring is ~ 7 eV, which decreases down to ~ 2 eV for a cluster of 20 aromatic rings. Thus, we expect that much smaller $sp^2$ clusters of few aromatic rings or of some other $sp^2$ configuration of similar size are likely to be responsible for the observed blue PL. Our previously proposed structural model for GO takes into account both the larger $sp^2$ domains and also smaller $sp^2$ fragments which are responsible for the transport of carriers between the larger $sp^2$ domains[14]. In this view, the observed increase in the PL intensity without energy shift during the initial reduction treatments may be attributed to the increased concentration of such $sp^2$ fragments. Furthermore, the subsequent PL quenching with longer reduction may be the result of percolation among these $sp^2$ configurations, facilitating transport of excitons to non-radiative recombination sites.

To test this hypothesis, we investigated the electrical transport properties of individual sheet of GO progressively reduced by exposure to hydrazine. We have recently shown that electrical properties of GO are sensitively dependent on $sp^2$ carbon fraction and spatial distribution of the smaller domains, serving as an indirect probe for the structural information regarding the evolution of the $sp^2$ phase with reduction[14, 36]. Figure 3a shows the *I-V* characteristics of an identical GO device reduced at different levels. It can be seen that the low-bias current is suppressed in GO reduced for 2 min whereas dramatic increase in current is observed after 5 min of reduction. The conduction is dominated by tunneling for GO reduced for 2 min while hopping begins to contribute to conduction after longer reduction (> 5 min) (See our Ref. [36] for a complete study of transport in progressively reduced GO). The transfer characteristics of these devices further indicate that GO essentially remains an insulator after 2 min exposure to hydrazine vapor (Figure 3b). After 5 min of reduction, ambipolar field effect becomes observable albeit with low current. These trends observed in the electrical properties are in striking contrast to the optical properties presented in Figure 1a and b. The general trends in absorbance, PL intensity and electrical conductivity with reduction time are summarized in Figure 3c. It can be noticed that the initial rapid rise of the PL intensity is coupled with the absorbance behavior, while the gradual quenching can be correlated with gradual increase in the conductivity. That is, while the absorbance of GO thin films increases immediately after only 20 sec of exposure to hydrazine vapor, the electrical conductivity remains low.

These results are consistent with our hypothesis that at the initial stages of reduction process, the fraction of strongly localized $sp^2$ sites increases, thereby improving absorbance and PL intensity while the energetic coupling between these sites remain negligibly small. Recent photoconductivity study also indicate that e-h pairs can be photo-generated and dissociated under relatively large electric field (> 400 V/cm) in reduced GO films[39]. In the later stages of reduction process, interconnectivity of the localized $sp^2$ sites increases, thereby facilitating hopping of excitons to non-radiative recombination centers and consequently quenching PL. Based on these observations, we have refined our previous structural model for GO at different stages of reduction to explain the PL results as shown in Figure 4. A schematic representing the structure of as-prepared GO shows the ~ 3 nm $sp^2$ clusters (not drawn to scale), along with smaller $sp^2$ configurations dispersed in an insulating $sp^3$ matrix where a large fraction of carbon is bonded with oxygen (oxygens are represented by orange dots) is shown in Figure 4a. Initially, the concentration of smaller $sp^2$ domains is low, yielding moderate PL emission. In Figure 4b, the evolution of the smaller $sp^2$ domains due to the removal of some oxygen with reduction is shown. Based on our previous study[14], we have found that the dimensions of the larger $sp^2$ clusters do not change but that the transport is mediated by the increase in concentration and size of the smaller $sp^2$ domains. The formation of new isolated smaller domains consisting of few conjugated repeating units, as shown in Figure 4b, results in the enhancement of the PL intensity observed between 20 s – 3 min of reduction. In Figure 4c, further removal of oxygen with continued exposure to hydrazine vapor leads to percolation between the larger $sp^2$ clusters via growth of smaller $sp^2$ domains is shown. Based on simple particle-in-a-box argument, it can be readily surmised that the PL from the structure shown in Figure 4c is likely to be significantly quenched, as observed in Figures 1b. It has been suggested that the non-radiative lifetime of e-h pairs govern the experimental PL decay times[31]. The above argument is further supported by our time-resolve PL measurements (See Supplementary Information), which show shorter decay time for GO films reduced for more than 3 minutes. The band diagram corollary of the physical picture is represented in Figure 4d where the blue PL from excitation and recombination among the discrete energy levels of small $sp^2$ fragments is represented. The energy states are highly localized due to the large σ-σ* gap of the $sp^3$ matrix (not shown in schematics), which is expected to be on the order of ~ 6 eV (Ref. [23]). The large $sp^2$ clusters (~ 3 nm) have energy gaps that are lower than those of the smaller $sp^2$ fragments and largely determine the total electronic density of states (DOS) of the material. The excitation and recombination in discrete energy states and larger gap of the small fragments as shown in Figure 4d are likely responsible for the enhanced blue PL emission in slightly reduced GO.

In summary, we describe blue PL from chemically derived GO thin films deposited from thoroughly exfoliated suspensions. The presence of isolated $sp^2$ clusters within the carbon-oxygen $sp^3$ matrix leads to localization of electron-hole pairs, facilitating radiative recombination for small clusters. The PL intensity was found to vary with reduction treatment, which can be correlated to the evolution of very small $sp^2$ clusters. Our calculations suggest that $sp^2$ clusters of several conjugated repeating units yield band gaps consistent with blue emission. The overall intensity of PL was moderate due to the lower total cross-section of emitting "molecular" $sp^2$ domains centers relative to non-radiative recombination sites. The possibility of engineering desired "molecular" $sp^2$ structures in graphene via controlled oxidation to achieve highly efficient and tunable PL will be important towards exploitation of blue PL for opto-electronics applications. These results call for further experiments based on temperature dependence of PL for better clarification of the electronic processes occurring in GO.

*Acknowledgements*
This research funded by the NSF CAREE Award (ECS 0543867). We acknowledge financial support from the Rutgers University Academic Excellence Fund and Institute for Advanced Materials, Devices and Nanotechnology (IAMDN). The authors would also like to thank Mr. Wei-Jung Lai at the Center for Condensed Matter Sciences, NTU for assistance with PLE measurements.

Figures

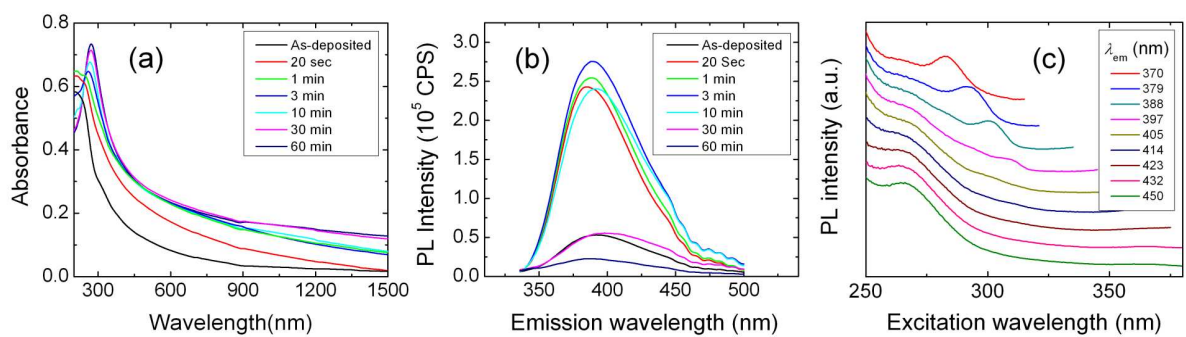

**Figure 1.** (a) Absorbance and (b and c) photoluminescence spectra of progressively reduced GO thin films. The total time of exposure to hydrazine is noted in the legend. The photoluminescence spectra in (b) were obtained for excitation at 325 nm. The PL excitation spectra in (c) were obtained for different wavelengths of the emission spectrum ranging from 370 to 450 nm.

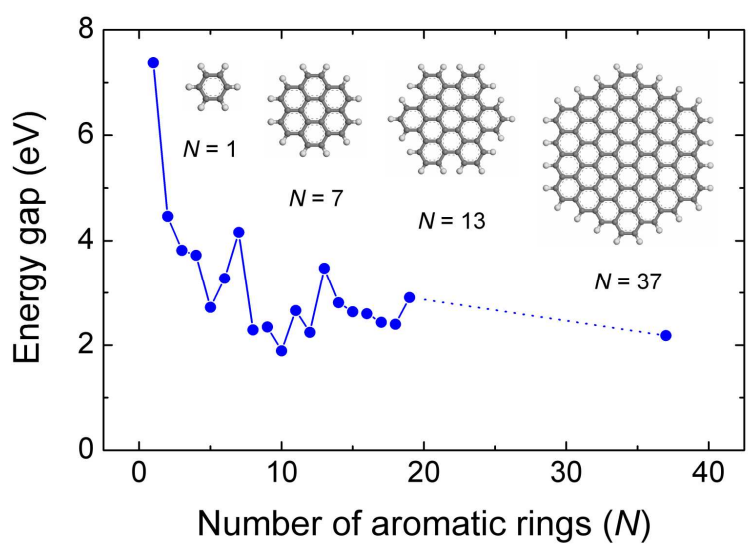

**Figure 2.** Energy gap of π-π* transitions calculated based on density functional theory (DFT) as a function of the number of fused aromatic rings ($N$). The inset shows the structures of graphene molecules used for calculation.

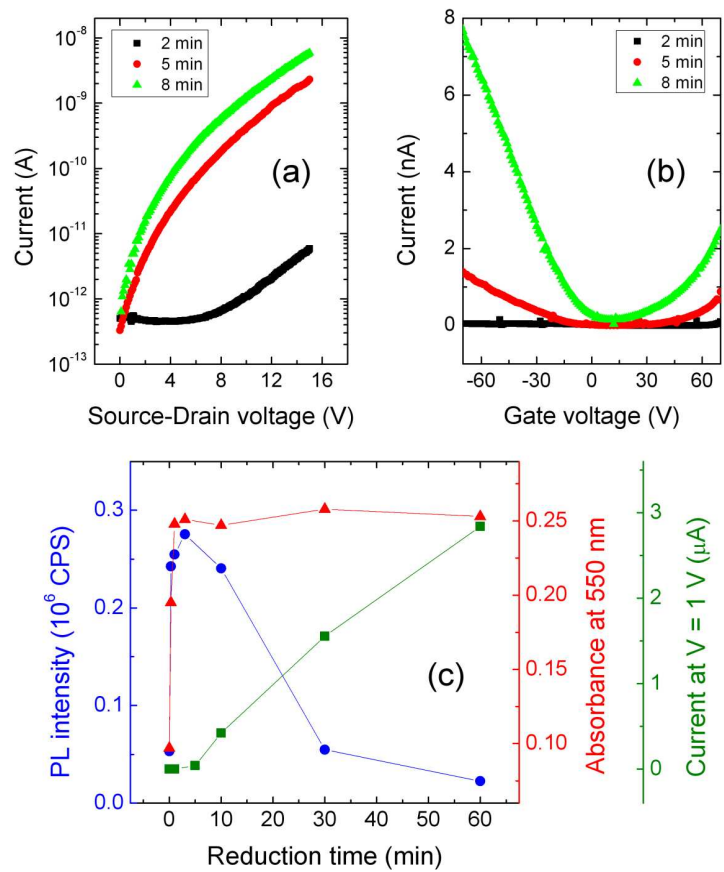

**Figure 3.** (a) *I-V* and (b) transfer characteristics of individual GO sheet devices at different stages of reduction. The total time of exposure to hydrazine is noted in the legend. (c) A summary plot showing the maximum photoluminscence intensity, absorbance at 550 nm, and current at 1 V for GO thin film as a function of reduction time.

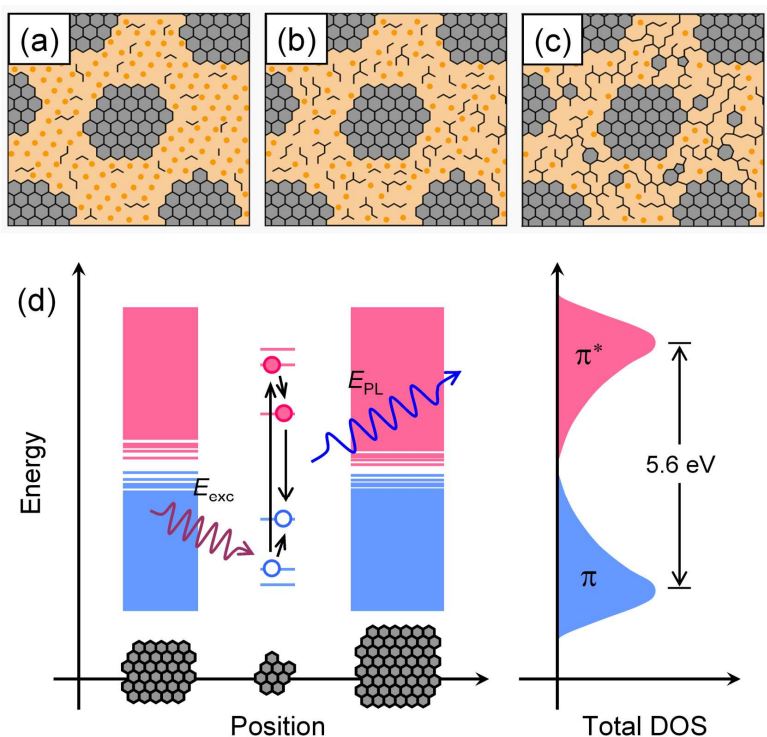

**Figure 4.** (a-c) Structural models of GO at different stages of reduction. The larger $sp^2$ clusters of aromatic rings are not drawn to scale. The smaller $sp^2$ domains indicated by zigzag lines do not necessarily correspond to any specific structure (such as olefinic chains for example) but to small and localized $sp^2$ configurations that act as the luminescence centers. The PL intensity is relatively weak for (a) as-synthesized GO but increases with reduction due to (b) formation of additional small $sp^2$ domains between the larger clusters due to evolution of oxygen with reduction. After extensive reduction, the smaller $sp^2$ domains create (c) percolating pathways among the larger clusters. (d) Representative band structure of GO. The energy levels are quantized with large energy gap for small fragments due to confinement. A photogenerated e-h pair recombining radiatively is depicted.

# Supplementary Information for
# "Blue photoluminescence from chemically derived graphene oxide"

Goki Eda[1], Yun-Yue Lin[3], Cecilia Mattevi[1], Hisato Yamaguchi[2], Hsin-An Chen[3], I-Sheng Chen[3], Chun-Wei Chen[3*], and Manish Chhowalla[1,2†]

[1] *Department of Materials, Imperial College, Exhibition Road, London SW7 2AZ, UK.*
[2] *Department of Materials Science and Engineering, Rutgers University*
*607 Taylor Road, Piscataway, NJ 08854, USA.*
[3] *Department of Materials Science and Engineering, National Taiwan University*
*No. 1, Sec. 4, Roosevelt Road, Taipei 10617, Taiwan.*

## 1. Preparation of highly exfoliated GO suspensions

Graphene oxide (GO) was synthesized by modified Hummers method[S1]. Suspension of GO were prepared after 7 cycles of purification by dilution and sedimentation. GO suspensions were then centrifuged at 5000 rpm for 90 min. The supernatant of this suspension was further centrifuged at 8000 rpm for 20 min. We found that almost all the GO sheets in the supernatant are completely exfoliated. The high degree of exfoliation has been verified by depositing ~ 1 monolayer thin films from the suspensions using vacuum filtration method (Ref. [S2]) and observing the film morphology under optical microscope. On $SiO_2$/Si substrates, monolayer of GO is readily distinguishable from multi-layers. The optical microscope images in Figure S1 show that films prepared from highly exfoliated GO suspensions are highly uniform whereas those made from suspensions prior to centrifugation contains thicker multi-layers. Thick particles also appear to cause the formation of pin-holes in the films.

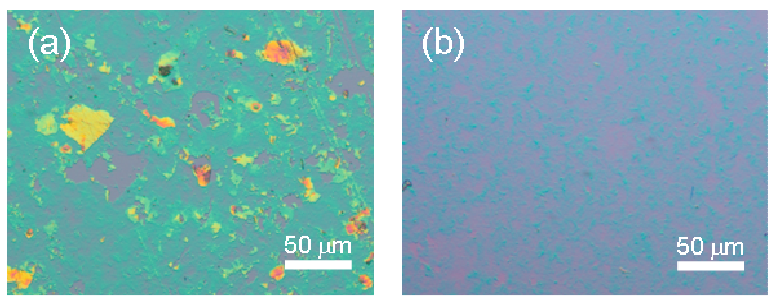

Fig S1: Optical microscope image of GO thin films deposited from GO suspension (a) before and (b) after centrifugation 8000 rpm for 20 min.

## 2. Optical measurements

UV–visible absorption spectra were obtained using a Jasco V570 UV/Vis/NIR spectrophotometer. .The PL spectra were obtained by exciting the samples using a continuous wave He-Cd laser 325 nm and the emission spectra were analyzed with a Jobin–Yvon TRIAX 0.55 m monochromator and detected by a photomultiplier tube and standard photocounting electronics. Time-resolved photoluminescence spectroscopy was performed with a time-correlated single photon counting (TCSPC) module (Picoharp 300, PicoQuant) with a temporal resolution 4ps. A pulse laser (372 nm) with an average power of 1 mW operating at 40 MHz with duration of 70 ps was used for excitation. Photoluminescence excitation (PLE spectroscopy was measured using a FluoroLog®-3 spectrofluorometer (Jobin-Yvon).

## 3. Electrical measurements

For electrical measurements of individual GO sheets, GO sheets were deposited randomly on to $SiO_2$/Si substrates by dip-coating. Standard e-beam lithography followed by lift-off process was used to define electrodes on the GO sheets. The electrode metals (Cr/Au) were thermally evaporated prior to lift-off. The electrical measurements were conducted using Agilent hp4140b parameter analyzer. Si substrate was used as the back gate electrode. The measurements for the data presented in Figure 3a and 3b were made in vacuum. The channel length and widths were 2.8 μm and 14.8 μm, respectively. Details of the electrical properties of reduced individual sheet of GO can be found in Ref. [S3-8]. For thin films, Au electrodes were thermally evaporated with a shadow mask. The electrical measurements on thin films presented in Figure 3c were conducted in air. The channel length and the width of the gap cells on films were 200 and 400 μm, respectively. Upon reduction, the sheet resistance decreased from ~$10^{11}$ to 3 x $10^5$ Ω/sq. See Ref. [S9] for detailed work.

## 4. Energy gap calculations

Electronic structure calculations were performed based on density functional theory (DFT) by using DMol$^3$ code. All electrons for core treatment and DNP for numerical basis set were used during calculations. The density functional was treated by the local density approximation (LDA) with exchange-correlation potential. We also used Gaussian to optimize the electronic structures by DFT using the B3LYP hybrid functional including 6-31g (d) basis. The ground state energy was calculated by DFT and excited state energy was calculated by TD-DFT, respectively. Orbital calculations were selected in both DMol$^3$ and Gaussian energy calculations.

The calculation result shows that the band gap energy decreases with the quantity of aromatic ring increasing in the three calculation strategies in general. If the molecule has a three-fold or six-fold symmetry (3, 7, 13, 16 and 19), the band gap energy will be higher. But it is not clear that some of the molecules have abnormal low band gap energy (5, 8 and 12).

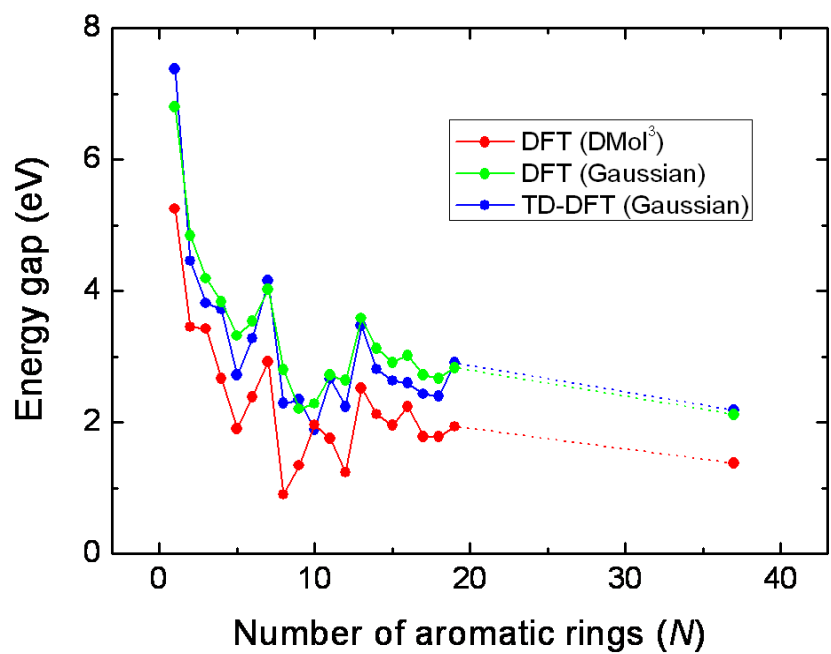

Fig S2: Energy gap of fused aromatic ring clusters calculated using three different methods.

## 5. Absorption of non-bonding state

In as-deposited GO thin films, we observed a weak absorption band near 300 ~ 320 nm which can be attributed to n-π* transitions of C=O bonds[S10]. This absorption peak was found to almost disappear immediately after reduction as shown in Fig S3. This observation is consistent with the removal of oxygen functional groups with reduction.

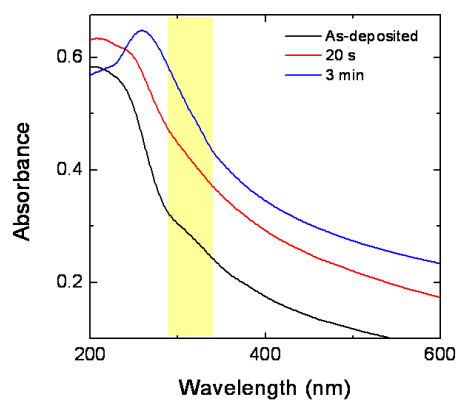

Fig S3: Magnified absorbance spectra from Fig 1a. The spectra for other exposure times are not shown for clarity.

## 5. PL characteristics of GO suspensions

We studied the PL characteristics of GO suspensions before and after the centrifugation which remove thick multi-layered GO sheets. We observed that the PL spectra of the suspensions were significantly different as shown in Figure S4. It should be noted that the PL spectrum obtained from GO suspensions prior to centrifugation resemble the spectra reported by Liu *et al*.[S11] and Zhengtang *et al*.[S12] The details of this study will be reported elsewhere.

The PL excitation spectrum of highly exfoliated GO suspension shown in Figure S5 exhibit excitonic absorption feature around 320 nm similar to the result of thin films presented in Figure 1c.

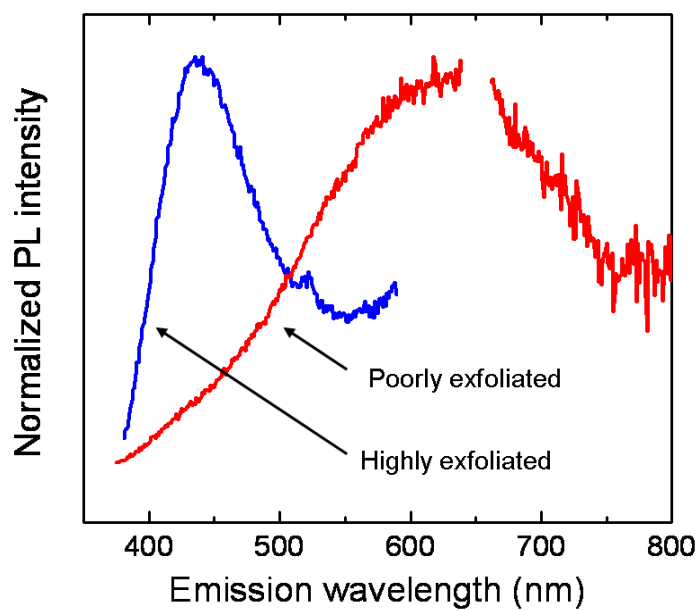

Fig S4: PL spectra of GO suspensions before and after centrifugation. The excitation wavelength was 320 nm.

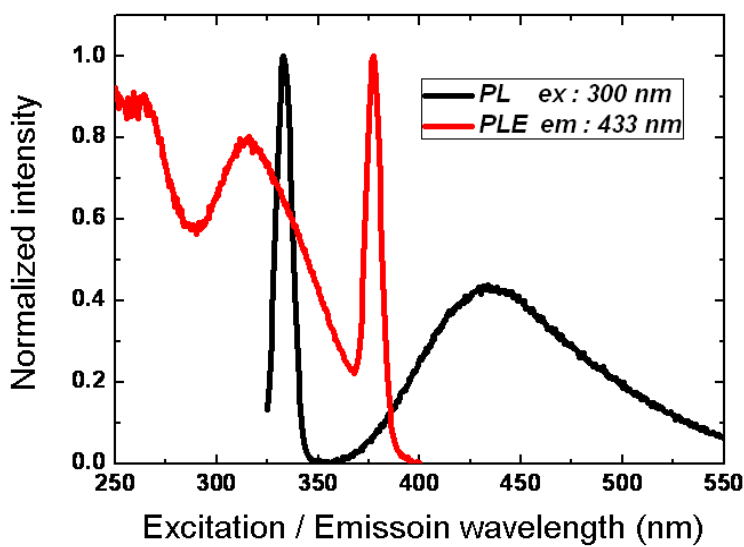

Fig S5: PL emission and excitation spectra of highly exfoliated GO suspensions. The spectra were obtained at excitation and emission wavelength of 300 nm and 433 nm, respectively. The sharp intense peaks (~ 330 nm for emission and ~ 375 nm for excitation spectrum) are the water Raman peaks.

## 6. Annealing effect on the PL characteristics of GO thin films

In order to verify that the PL from GO thin films is not artifacts introduced during deposition process or via exposure to hydrazine, we studied the PL of GO films annealed at different temperatures (Figure S6). The PL peak was weaker but still present after annealing at 200 °C.

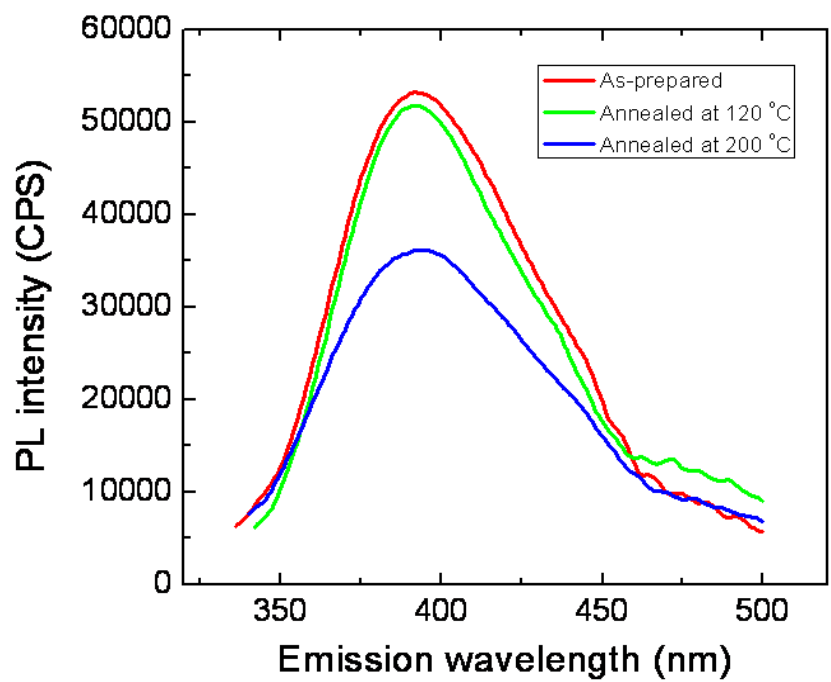

Fig S6: PL spectra of GO thin films annealed at different temperatures.

## 7. Time-resolved photoluminescence

Time-resolved PL (TRPL) was performed to obtain exciton lifetimes. It can be seen from Figure S7 that the exciton lifetimes increase with hydrazine vapor exposure time up to 3 minutes. This is consistent with the observed PL intensity increase in Figures 1b and 1c. As-prepared GO is insulating because the $sp^2$ clusters are interrupted by oxygen functional groups, which limits the density of e-h pairs created by excitation. Additional $sp^2$ clusters begin to form upon longer exposure to hydrazine vapor[S13], which leads to increased number of localized states allowing radiative recombination to occur. The exciton life time was found to be weakly dependent on the reduction time in this stage of reduction. Even longer hydrazine vapor exposure times leads to percolation of *sp*² cluster, facilitating delocalization of e-h pair. Reduction is also expected to produce defect such as vacancies[S14]. Increased number of defects and larger delocalizaion length in reduced GO result in higher probability of non-radiative recombination and therefore faster decay of PL. Decreasing exciton lifetime is demonstrated in Figure S6b. Figure S8 shows that non-radiative recombination can occur by tunneling or hopping of e-h pairs into defect sites where deep trap levels are present.

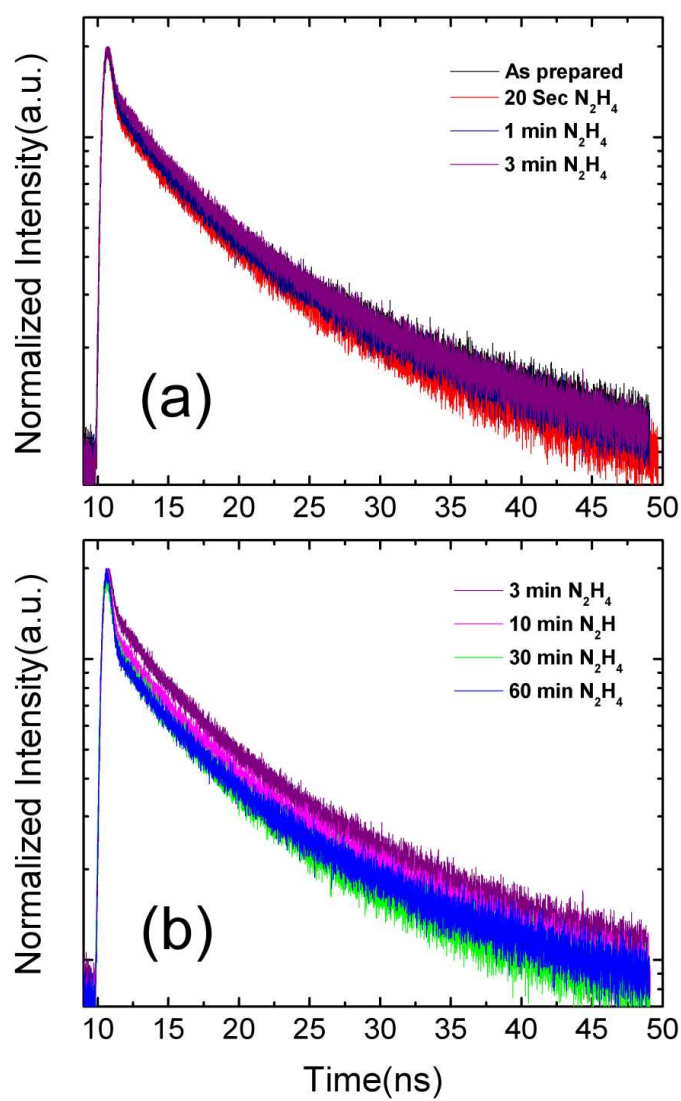

Fig S7: Time-resolved PL of GO with different degree of reduction. The PL life time (a) increases gradually over initial 3 min of hydrazine vapor exposure but then (b) decreases for longer exposure times.

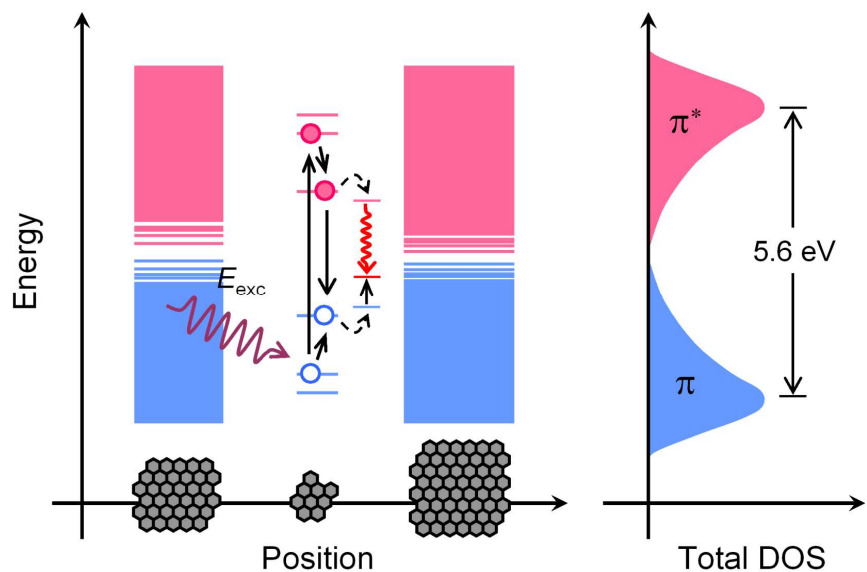

Fig S8: Band structure model of GO. The density of states (DOS) of $sp^2$ clusters depend on their size and shape. The energy levels for small clusters are quantized due to confinement effect. Non-radiative recombination can occur when e-h pairs tunnel or hop to a defect site and decay into deep trap levels.